\documentstyle[12pt]{article}

\newcommand{\ux}{\underline{x}}
\newcommand{\hlp}{\hat{L}_{\rm part}}

\newcommand{\ue}{\underline{\eta}}
\newcommand{\ha}{\hat{A}}

\begin{document}

\title{Deformed Chern-Simons interaction for nonrelativistic point
particles}
\author{P.C. Stichel\\
An der Krebskuhle 21\\D-33619 Bielefeld\thanks{E-mail: pstichel@gmx.de}}

\date{ }

\maketitle

\begin{abstract}
We deform the interaction between nonrelativistic point particles on a plane and a
Chern-Simons field to obtain an action invariant with respect to
time-dependent area-preserving diffeomorphisms. The deformed and
undeformed Lagrangians are connected by a point transformation leading
to a classical Seiberg-Witten map between the corresponding gauge
fields. The Schroedinger equation derived by means of Moyal-Weyl
quantization from the effective two-particle interaction exhibits

-- a singular metric, leading to a splitting of the plane into an
interior (bag-) and an exterior   region,

-- a singular potential (quantum correction) with singularities located
at the origin and at the    edge of the bag.

\noindent
We list some properties of the solutions of the radial Schroedinger
equation.
\end{abstract}


\section{Introduction}

Two-dimensional incompressible fluids, in particular quantum-hall fluids
(QHF's) are well known to be invariant with respect to time-independent
area-preserving diffeomorphisms (cp.[1]). In a particle picture a QHF is
usually described by neglecting the kinetic energy compared to the
magnetic field term leading to noncommutative geometry and a reduced
phase space (cp.[2]). In the present paper we generalize this
picture by allowing
\begin{description}
\item{--} the particles to move in full phase space,

\item{-} the area-preserving diffeomorphisms to become time-dependent.
\end{description}

\noindent
To do this we consider a deformation of the interaction of
nonrelativistic charged point particles on a plane coupled to a Chern-Simons (CS)
field such that

\smallskip
\noindent
\begin{description}
\item{--} the deformed action is invariant with respect to time-dependent
area-preserving diffeomorphisms $\nu_{2,t}$,

\item{--} the deformed and undeformed particle Lagrangians are connected by a
point transformation leading to the classical analogue of a
Seiberg-Witten (SW) map between the deformed and undeformed gauge
fields.
\end{description}

There is a second reason for studying such a model. A theory, unifying
translational- and $U(1)$-gauge invariance in 2d contains nine independent
gauge fields: six dreibein components and three $U(1)$-gauge fields [3].
Our model shows that the restriction of the group of local translations
to its subgroup $\nu_{2,t}$ reduces the number of independent gauge fields to
three as the dreibeins are built from the deformed U(1)-gauge fields.

\noindent
We begin by  constructing the deformed Lagrangian, consider the
equations of motion (EOM) and discuss the properties of the deformed
gauge fields. After gauge fixing we solve the nonlinear Gauss-constraint
for the two-particle problem and quantize it by means of the Moyal-Weyl
prescription. We discuss the structure of the resulting radial
Schroedinger equation and list some properties of its solutions.
We conclude with some final remarks.

\section{Deformed Lagrangian}

Infinitesimal elements of $\nu_{2,t}$ are defined by\footnote{Such transformations are used also in [4].}
\begin{equation}
\delta x_i = - \theta \epsilon_{ij} \partial_j \Lambda (\ux,t) \ \ , \ \ \ i,j =1,2             
\end{equation}
at fixed time $t$, where $\Lambda$ is an infinitesimal gauge function and $\theta$ is
a finite deformation parameter. The corresponding change of a 
field $f(\underline{x},t)$ is defined by
\begin{equation}
\delta_0 f (\ux,t): = f^\prime (\ux,t) - f(\ux,t)
\end{equation}
or, if we include the coordinate change, we define
\begin{equation}
\delta f (\ux,t) : = f^\prime (\ux^\prime,t) - f(\ux,t)
\end{equation}
Now, we want to deform
\begin{equation}
L_{\rm{part}} : = \frac{1}{2} \dot{x}_i^2 + e(A_i (\ux,t) \dot{x}_i + A_0 (\ux,t)) 
\end{equation}
in such a way that $\hat{L}_{part}$ becomes quasi-invariant with respect to (1) where 
$\hat{A}_\mu$ transform as\footnote{Deformed quantities are marked by a hat.}

\begin{equation}
\delta \hat{A}_\mu (\ux,t) = \partial_\mu \Lambda (\ux,t)
\end{equation}
or, equivalently 
\begin{equation}
\delta_0 \hat{A}_\mu (\ux,t) = \partial_\mu \Lambda (\ux,t) + \theta \epsilon_{ij} \partial_i \hat{A}_\mu \partial_j \Lambda 
\end{equation}
ie. the gauge transformations of the $\hat{A}_\mu$ fields mix local $U(1)$-transformations
with space transformations (1) (cp.~[5])\footnote{The idea of mixing coordinate transformations with gauge transformations has been developed in [6] and extended to noncommutative space in [7].}.

In order to deform the first term in (4) we define invariant
coordinates $\eta_i$  (cp.~[8])
\begin{equation}
\eta_i (\ux,t) : = x_i + \theta \epsilon_{ij} \hat{A}_j (\ux,t).
\end{equation}           
Obviously, we have
\begin{equation}
\delta \eta_i = 0\ .
\end{equation}
Then, by means of the invariant velocity $\xi_i$
\begin{equation}
\xi_i : = \frac{d}{dt} \eta_i (\ux,t),
\end{equation}
we define $\hat{E}_{kin}$
\begin{equation}
\hat{E}_{kin} : = \frac{1}{2} \xi^2_i\ .
\end{equation}                                  
The deformation of the interaction term in (4) is more involved. It is
given by
\begin{equation}
e(A_i \dot{x}_i + A_0) \to e (\hat{A}_i \dot{x}_i + \hat{A}_0) + \frac{e\theta}{2} \epsilon_{ij} \hat{A}_i \frac{d}{dt} \hat{A}_j = : \hat{L}_{int},
\end{equation}
where an additional CS-term is needed because $\delta  \dot{x}_i \not= 0$ and 
because we want
 $\hat{L}_{int}$ to be quasi-invariant:
\begin{equation}
\delta \hat{L}_{int} = e \frac{d}{dt} ( \Lambda - \frac{\theta}{2} \epsilon_{ij} \hat{A}_i \partial_j \Lambda).
\end{equation}
To do this it is advantageous to replace $\hat{L}_{part}$ by its first-order
form
\begin{eqnarray}
\hat{L}_{\rm part} &=& \dot{x}_k (\xi_k + e \hat{A}_k + \theta \epsilon_{ij} (\xi_i + \frac{e}{2} \hat{A}_i)\partial_k \hat{A}_j) - \frac{1}{2} \xi_i \xi_i\nonumber\\
& & + \theta \epsilon_{ij} (\partial_t \hat{A}_j) (\xi_i + \frac{e}{2} \hat{A}_i) + e\hat{A}_0.
\end{eqnarray}

\noindent
By varying the action $\hat{S}_{part}$ with respect to $\xi_i$ and $x_i$ we get (9) as a
constraint
\begin{equation}
\xi_i = \dot{x}_i + \theta \epsilon_{ij} \frac{d}{dt} \hat{A}_j
\end{equation}           
and the invariant nonlinear Lorentz-force equation
\begin{equation}
\dot{\xi}_k = \frac{e}{1+\theta \hat{F}} (\epsilon_{kl} \xi_\ell \hat{F} + \hat{F}_{ko})
\end{equation}
with the invariant field strength $\hat{F}_{\mu\nu}$
\begin{equation}
\hat{F}_{\mu\nu} : = \partial_\mu \hat{A}_\nu - \partial_\nu \hat{A}_\mu + \theta \epsilon_{ik} (\partial_i \hat{A}_\mu) \partial_k \hat{A}_\nu
\end{equation}
and
\begin{equation}
\hat{F}_{ij} = \epsilon_{ij} \hat{F}\ . 
\end{equation}

We note that in the particular case of a constant external magnetic
field B, ie. for $\hat{F} = B$, (14) and (15) are equivalent to the EOM given by
Duval and Horvathy [9] for a particle which possesses a nonvanishing second
central charge $k$ of the planar Galilei group [10] with  $e k = -\theta$.

\noindent
Finally, the CS-interaction of the $\hat{A}_\mu$ field invariant with respect to the gauge
transformations (5), is given by [5]
\begin{equation}
\hat{L}_{CS} = \frac{\kappa}{2} \int d^2 x \epsilon^{\mu\nu\rho} \hat{A}_\mu (\partial_\nu \hat{A}_\rho  + \frac{\theta}{3} \epsilon_{ik} (\partial_i \hat{A}_\nu) (\partial_k \hat{A}_\rho)).
\end{equation}

\section{Deformed gauge fields}

Usually the invariant velocity $\xi_i$ is defined in terms of nonrelativistic
dreibein fields $E_\mu^\nu$ $(\mu,\nu = 0,1,2)$\footnote{The space indices can be taken equivalently as lower or upper
indices.} [8]
\begin{equation}
\xi_i = E^i_k \dot{x}_k + E^i_0\ .
\end{equation}
Comparing (19) with (14) and considering time as fixed leads to dreibeins
expressed in terms of gauge fields $\hat{A}_k$
\begin{equation}
E_\mu^a : = \theta \epsilon_{ak} \partial_\mu \hat{A}_k  + \delta_{a\mu}
\end{equation}
$$
E^0_0 = 1\ \ \ \mbox{and} \ \ \ E^0_k = 0\ ,
$$
which, due to (5), transform covariantly with respect to $\nu_{2,t}$.
Note that in the case of arbitrary local translations as an invariance
group, dreibeins and the $\hat{A}_\mu$ are independent of each other [3]. Only the
restriction to the subgroup $\nu_{2,t}$ allows the relation (20).

\noindent
From the transformation law (6) we infer that the $\hat{A}_\mu$ are the gauge
fields of the classical limit of a noncommutative $U(1)$-gauge theory.
This raises the question of a possible Seiberg-Witten (SW) map [9]
between the deformed and undeformed gauge fields $\hat{A}_\mu$ and $A_\mu$.
For this we consider the point transformation
\begin{equation}
x_i \to \eta_i (\ux,t)
\end{equation}
and redefine the gauge fields
\begin{equation}
\hat{A}_\mu (\ux,t) \to A_\mu (\ue,t)
\end{equation}
so that
\begin{equation}
\hlp (\hat{A}_\mu (\ux,t), \dot{x}_i) = L_{\rm part} (A_\mu (\ue,t), \dot{\eta}_i)
\end{equation}
(21)-(23) defines the classical analogue of an inverse SW-map between
gauge fields on a commutative space.\footnote{A different definition of such a classical SW-map has been given in
[13].}

\bigskip
\noindent
This inverse SW-map may be given explicitly in terms of inverse
dreibeins $\{e^\nu_\mu\}$  
\begin{equation}
A_\nu (\underline{\eta} (\ux ,t),t) = \frac{1}{2} \hat{A}_\mu (\ux , t) (\delta^\mu_\nu + e^\mu_\nu (\ux ,t))
\end{equation}
with, according to (20),
\begin{equation}
e^i_j = \frac{\delta_{ij} - \theta\epsilon_{ik} \partial_k \hat{A}_j}{1+\theta \hat{F}},
\end{equation}
$$
e^i_0 = -\theta \epsilon_{jk}\partial_t \hat{A}_k e^i_j,
$$
$$
e^0_0 = 1\ \ \ \mbox{and} \ \ \ e^0_i = 0\ .
$$
Solving (24) for $\hat{A}_\mu$ to leading order in $\theta$ we obtain
\begin{equation}
\hat{A}_\mu = A_\mu - \frac{\theta}{2} \epsilon_{ik} A_i (\partial_k A_\mu + F_{k\mu}) + 0 (\theta^2)
\end{equation}
in agreement with [12].

\section{Gauge fixing and residual symmetry}

$\hat{A}_0$ is a Lagrange multiplier whose variation in the total action
\begin{equation}
\hat{S} = \hat{S}_{\rm part} + \hat{S}_{\rm field}
\end{equation}
leads to the Gauss-constraint
\begin{equation}
\hat{F} (\ux,t) = - \frac{1}{\kappa} \sum^N_{\alpha=1} e_\alpha \delta (\ux - \ux_\alpha).
\end{equation}
Eq. (28) can be integrated once
\begin{equation}
\hat{A}_k + \frac{\theta}{2} \epsilon_{\ell S} \ha_\ell \partial_k \ha_S = - \frac{1}{2\pi \kappa} \sum_\alpha e_\alpha \partial_k \phi (\ux - \ux_\alpha) + \partial_k \lambda (\ux,t)
\end{equation}
where
\begin{equation}
\phi(\ux):= \arctan \frac{x_2}{x_1}
\end{equation}
is a singular gauge function  which has to be regularized (cp.~[11])
and $\lambda$ is an arbitrary gauge function to be determined by fixing
the asymptotic behaviour of $\hat{A}_\mu$. For that we follow closely the
procedure described in [11]:

\smallskip
\noindent
\begin{description}
\item{i)} We decompose 
\begin{equation}
\ha_\mu = \tilde{A}_\mu + \hat{A}^{as}_\mu 
\end{equation}
with                   
\begin{equation}
\tilde{A}_\mu {\buildrel r\to\infty \over \to} 0 (r^{-1}).
\end{equation}
In order to fulfill (32) the $\tilde{A}_i$ should be chosen as solutions of
(29)$_{\lambda=0}$.

\item{ii)} We require the asymptotically Euclidean metric leading to
\begin{equation}
\ha_j^{as} = \frac{1}{\theta} \epsilon_{jk} a_k (t)
\end{equation}
ie. the $\hat{A}^{as}_j$ transform covariantly with respect to translations, local
in time (residual symmetry). This requirement fixes
also $\hat{A}^{as}_0$.
Thus our procedure fixes $\lambda$ (gauge fixing).

\item{iii)} We redefine the Lagrangian in terms of the new variables $\{\tilde{A}_\mu, a_i(t)\}$ such that the solutions of the Euler-Lagrange equations minimize
the new action.
\end{description}

\section{The classical two-particle problem}

We consider two identical particles of charge $e$. Applying the Legendre
transformation to our Lagrangian and using the Gauss-constraint (28) we
obtain
\begin{equation}
H = \frac{1}{2} \sum^2_{\alpha = 1} \xi^2_{i,\alpha}
\end{equation}
with the constraint
\begin{equation}
\sum^2_{\alpha=1} \xi_{i,\alpha} = 0
\end{equation}
arising from the variation of $\dot{a}_i (t)$ in the redefined action.
In order to express the $\xi_{i,\alpha}$ in terms of  canonical variable $\{\ux_\alpha , \underline{p}_\alpha \}$
we need the $\tilde{A}_k$ at the particle positions $\ux = \ux_{\buildrel 1\over 2}$ to be solutions of
\begin{equation}
\tilde{A}_k + \frac{\theta}{2} \epsilon_{\ell S} \tilde{A}_\ell \partial_k \tilde{A}_S = \frac{e}{2\pi \kappa} \epsilon_{k\ell} \sum^2_{\alpha =1} \left( \frac{(x-x_\alpha)_\ell}{|\ux - \ux_\alpha |^2}\right)_{reg}
\end{equation}
We find in obvious notation
\begin{equation}
\tilde{A}_{k, {\buildrel 1\over 2}} = \pm \epsilon_{k\ell} (x_1 - x_2)_\ell \chi (|\ux_1 - \ux_2 |)
\end{equation}
with
\begin{equation}
\chi(r) : = \frac{1}{\theta} \left( 1 - \left( 1 - \frac{\tilde{\theta}}{r^2}
\right)^{1/2}\right)
\end{equation}
where
\begin{equation}
\tilde{\theta} : = \frac{e\theta}{\pi \kappa}\ .
\end{equation}
With (37) and the position and momentum variables for the relative motion  
\begin{equation}
\ux : = \ux_1 - \ux_2 \ , \qquad \underline{p} = \frac{1}{2} (\underline{p}_1 - \underline{p_2})
\end{equation}
we obtain by means of a straightforward computation the two-particle
Hamiltonian $H$ in terms of canonical variables 
\begin{equation}
H = \left(p_k - \frac{e^2}{2\pi \kappa} \epsilon_{k\ell} \frac{x_\ell}{r^2}\right) \left( p_{k^\prime} - \frac{e^2}{2\pi \kappa} \epsilon_{k^\prime \ell^\prime} \frac{x_{\ell^\prime}}{r^2} \right) g^{k k^\prime}
\end{equation}
with the inverse metric tensor $g^{k k^\prime}$ given by
\begin{equation}
g^{kk^\prime} : = ( 1- \tilde{\theta}/r^2)^{-1} \delta_{kk^\prime} - \frac{\tilde{\theta} (2-\tilde{\theta}/r^2)}{r^2- \tilde{\theta}
} \frac{x_k x_{k^\prime}}{r^2}.
\end{equation}
In plane spherical coordinates (41) reads
\begin{equation}
H = p^2_r (1 - \tilde{\theta}/r^2) + \frac{(\ell + \frac{e^2}{2\pi \kappa})^2}{r^2 - \tilde{\theta}} 
\end{equation}
where $\ell$   is the canonical angular momentum
\begin{equation}
\ell : = \epsilon_{ik} x_i p_k.
\end{equation}
From (43), respectively (41,42), we conclude that $H$ is singular at $r^2 = \tilde{\theta} = : r^2_0$ 
if $\tilde{\theta} > 0$ and so that $E \stackrel{<}{>} 0$ for $r \stackrel{<}{>} r_0$. Thus we conclude that we have
no communication between the interior $(r < r_0)$ and the exterior $( r > r_0)$ space regions. We have a geometric bag determined by the singularity of our dynamically generated metric.

\section{The two-particle Schr\"odinger equation}

By applying the Moyal-Weyl quantization procedure (cp.~[14] eq.~(3.7)) to $H$ given in terms of
Cartesian variables (eq. (41)) we obtain the radial Schr\"odinger equation 
\begin{equation}
\left( - \frac{\hbar^2}{r} \partial_r r (1- \tilde{\theta}/r^2) \partial_r + \frac{\overline{m}^2}{r^2 - \tilde{\theta}} + V(r) - E\right) \varphi_{\overline{m}} (r) = 0 
\end{equation}
with a singular potential $V(r)$
\begin{equation}
V(r) : = - \frac{\hbar^2 \tilde{\theta}}{2} \left( \frac{1}{r^2 - \tilde{\theta})^2} - \frac{1}{r^4}\right) 
\end{equation}
and a fractional (anyonic) angular momentum 
\begin{equation}
\overline{m} : = m + \frac{e^2}{2\pi \kappa} \ , \qquad m \in {\bf Z}\ .
\end{equation}
Let us list some properties  of the solutions of (45):

\smallskip
\noindent
\begin{description}
\item{i)} For $r\to r_0$  we obtain a behaviour which is typical for the singularity
there (cp.~[15])
\begin{eqnarray}
\varphi_{\overline{m}}(r) \simeq \left\{ \begin{array}{ll} 
C(\tilde{\theta}-r^2)^{1/4} \exp \left( \frac{-(\tilde{\theta}/2)^{1/2}}{(\tilde{\theta}-r^2)^{1/2}}\right) & \mbox{if}\ \ r < r_0\cr
(r^2 - \tilde{\theta})^{1/4} \left( A \cos \left( \frac{(\tilde{\theta}/2)^{1/2}}{(r^2 -\tilde{\theta})^{1/2}}\right) \right. + &  \cr
\left. + B \sin \left( \frac{(\tilde{\theta}/2)^{1/2}}{(r^2-\tilde{\theta})^{1/2}}\right) \right) & \mbox{if} \ \ r > r_0 \end{array} \right.
\end{eqnarray}
With the required continuity of the partial radial current $j_{\overline{m}}$ at
$r=r_0$ we infer from (48) $_{r<r_0}$ that
\begin{equation}
j_{\overline{m}} (r_0) = 0\ .
\end{equation}
Therefore $A/B$ in (48) must be a real number.

\noindent
Due to (49) we have no communication between the interior (bag-) and the
exterior region also in the quantum case. In particular, a scattering
wave ariving from the exterior region is totally
reflected at the edge of the bag. Thus the  bag acts like a white hole.

\item{ii)} For $r\to 0$ we obtain 
\begin{equation}
\varphi_{\overline{m}} /r) \simeq (r^2)^{S_\pm}\ , \qquad S_\pm : = \frac{1}{2} \left( 1 \pm \frac{1}{\sqrt{2}} \right)
\end{equation}
ie. all solutions are regular at the origin (cp.~[15]).

\item{iii)} From (48) and (50) we infer that all solutions  are square
integrable within the bag region. As we have no additional boundary
condition at hand which would determine a discrete spectrum, we expect
the spectrum to be continuous. Such a situation is characteristic for
singular potentials (cp.~[15]).
\end{description}

\section{Final remarks}

We have shown that a deformed interaction between charged point
particles and a CS-field, made invariant with respect to time-dependent
area-preserving diffeomorphisms, leads to a two-particle Schroedinger
equation of a highly singular nature. We have obtained some properties of its
solutions  but a complete discussion of its solution structure is still
lacking.

\noindent
Work on the continuum  generalization and the inclusion of an external
magnetic field is in progress.

\bigskip
\noindent
{\large{\bf Acknowledgements}}

\medskip
\noindent
I'm grateful to J. Lukierski and W. Zakrzewski for valuable comments. I thank R.~Jackiw for pointing out the references [6] and [7].

\section{References}

\begin{description}
\item{[1]} S.~Bahcall, L.~Susskind, Int.~J.~Mod.~Phys. {\bf B5} (1991) 2735.

\item{[2]} G.V.~Dunne, R.~Jackiw, C.A.~Trugenberger, Phys.~Rev. {\bf D41} (1990) 661.\\
    G.V.~Dunne, R.~Jackiw, Nucl.~Phys. {\bf B 33C} (Proc. Suppl.) (1993) 114.

\item{[3]} J.~Lukierski, P.C.~Stichel, W.J.~Zakrzewski, Eur.~Phys.~J. {\bf C20} (2001)
759.

\item{[4]} B.~Freivogel, L.~Susskind, N.~Toumbas, hep-th/0108076.

\item{[5]} R.P.~Manvelyan, R.L.~Mkrtchyan, Phys.~Lett. {\bf B327} (1994) 47.

\item{[6]} R.~Jackiw, Phys.~Rev.~Lett.~{\bf 41} (1978) 1635.

\item{[7]} R.~Jackiw, S.-Y.~Pi, hep-th/0111122.

\item{[8]} J.~Madore, S.~Schraml, P.~Schupp, J.~Wess, Eur.~Phys.~J.~{\bf C16} (2000) 161.

\item{[9]} C.~Duval, P.A.~Horvathy, Phys.~Lett. {\bf B479} (2000) 284; J. Phys. {\bf A34} (2001) 10097.

\item{[10]} J.~Lukierski, P.C.~Stichel, W.J.~Zakrzewski, Ann.~Phys.~(N.Y.) {\bf 260}
(1997) 224.

\item{[11]} J.~Lukierski, P.C.~Stichel, W.J.~Zakrzewski, Phys.~Lett. {\bf B484} (2000)
315; Ann.~Phys.~(N.Y.) {\bf 288} (2001) 164.

\item{[12]} N.~Seiberg, E.~Witten, JHEP {\bf 9909} (1999) 032.

\item{[13]} B.~Jurco, P.~Schupp, J.~Wess, Nucl.~Phys. {\bf B584} (2000) 784.

\item{[14]} L.~Castellani, Class.~Quant.~Grav. {\bf 17} (2000) 3377.

\item{[15]} W.M.~Frank, D.J.~Land, R.M.~Spector, Rev.~Mod.~Phys. {\bf 43} (1971) 36.
\end{description}
\end{document}